\documentclass[a4paper]{jpconf}

\usepackage{graphicx}
\usepackage{color}
\usepackage{latexsym}
\usepackage{amsmath,amssymb}        
\usepackage[draft=false]{hyperref}

\begin{document}
\title{Toward restrictions on boson stars as black hole mimickers}

\author{F. S. Guzm\'an}

\address{Instituto de F\'{\i}sica y Matem\'{a}ticas, Universidad
              Michoacana de San Nicol\'as de Hidalgo.\\ Edificio C3, Cd.
              Universitaria, 58040 Morelia, Michoac\'{a}n,
              M\'{e}xico.}

\ead{guzman@ifm.umich.mx}

\begin{abstract}
The status of boson stars as black hole mimickers is presented among other mimickers. We focus on the analysis of the emission spectrum of a simple accretion disk model. We describe the free parameters that allow a boson star to become a black hole mimicker and present an example of a particular astrophysical case.
\end{abstract}


\section{Introduction}

One of the most important problems in relativistic astrophysics is that of the nature of black hole candidates (BHCs). It is commonly assumed that black hole solutions are correspond to BHCs. Nevertheless there are various other types of solutions of Einstein's equations sourced by different  matter models that provide similar behavior to high energy observations in the vicinity of BHCs. These solutions are known as black hole mimickers. Among those solutions there are wormholes \cite{Harko2009,Lemos2008}, gavastars \cite{Lemos2008,Visser2004,Rezzolla2007,Harko2009c}, brane world solutions \cite{Harko2008}, naked singularities \cite{Harko2010} and boson stars \cite{diego,diego-acc,Guzman2006,Guzman2009}. In fact, one important issue is to determine whether or not BHCs have an event horizon \cite{narayan}.

In order to estimate observable differences between black holes and mimickers is focused on the one hand to the search of gravitational wave fingerprints of the BHCs and their perturbative properties.
In the first place perturbations to Schwarzschild black holes are very well studied (for a recent review see e.g. \cite{Nagar2005}).  On the other hand, black hole mimickers' perturbations have not been studied as much, however some results are in turn, for instance, the stability of the whole set of wormhole solutions usually thought of as mimickers is not clear, instead the instability of wormhole solutions has been shown for basic wormhole solutions \cite{Hayward,GGS1,GGS2,GGS3}, destroying previous hope on the possibility that these are stable as shown in \cite{Armendariz} for particular types of perturbations. Concerning gravastars, stability has been explored and has been found that there are stability regions \cite{Visser2004,Carter2005,Rezzolla2008}. About boson star solutions the stability of the solutions has actually been exhaustively studied, for instance using perturbative methods \cite{Gleiser,scott}, catastrophe theory \cite{Catastro} and full non-linear numerical relativity, both in spherical symmetry  \cite{scott,SeidelSuen1990,Balakrishna1998} and full 3D \cite{Guzman2004}, and the stable branches of solutions are well known, and it has also been studied the fundamental quasinormal modes and tails of boson stars \cite{Lora2010}.

The study of boson stars has been pushed forward up to the binary boson star collision \cite{Lehner2007} and the non-linear evolution of perturbed boson stars \cite{ruxandra}, in both cases  considering the system as a source of gravitational waves. The study of gravitational wave signatures has also been studied in order to tell between a black hole and a gravastar \cite{Rezzolla2007}. Instead, in the case of wormholes e.g. simple solutions supported by a phantom scalar field would not stand the perturbative analysis and there is no hope for a study of a binary system,  because as shown in \cite{GGS2} the life time of these solutions is rather short and they should either collapse and form black holes or explode before they could merge.

On the other hand, a yet more common strategy to find observable differences among BHCs is associated to the study of the electromagnetic -observable- emission spectrum of geometrically thin optically thick disk models, since they are based on the study of time-like geodesics on a fixed background space-time and assume very simple conditions on the accreting material \cite{Harko2008,Harko2010,diego,diego-acc,Guzman2006,Guzman2009}.


\section{Boson stars}

Boson stars (BSs)  are spherically symmetric stationary solutions to Einstein's field equations minimally coupled to a complex scalar field  sourced by the stress energy tensor $T_{\mu \nu} = \frac{1}{2}[\partial_{\mu} \phi^{*} \partial_{\nu}\phi +
\partial_{\mu} \phi \partial_{\nu}\phi^{*}] -\frac{1}{2}g_{\mu \nu}
[\phi^{*,\alpha} \phi_{,\alpha} + V(|\phi|^2)]$, with $V(|\phi|) = m^2|\phi|^2 + \lambda |\phi|^4$ where $m$ is associated to the mass of the boson and $\lambda$ to the self-interaction of the boson system and the scalar field obeys the Klein-Gordon equation. The condition to find BSs solutions is that the scalar field has a harmonic dependence on time $\phi(r,t) = \phi_0(r) e^{-i \omega t}$,
where $r$ is the radial coordinate, $t$ the coordinate time and $\omega$ is a characteristic frequency that determines the central density of the solution \cite{Guzman2006}. This condition implies that the stress-energy tensor becomes time-independent, which in turn implies that the geometry
of the space-time is also time-independent. Using the rescaling
$\tilde{\phi}_0 = \sqrt{4\pi} \phi_0$,
$\tilde{r} = m r$,
$\tilde{t} = \omega t$,
$\tilde{\alpha} = \frac{m}{\omega}\alpha$ and
$\Lambda = \frac{2\lambda}{8 \pi m^2}$, 
one constructs the boson star solutions like those shown in Fig 1a. for various values of $\Lambda$. Each point of each curve represents a BS solution. Those to the left of the maxima are stable whereas those to the right are unstable \cite{Guzman2009RMF}.


\section{The accretion disk model}

The disk model around a given BHC is defined as follows. It corresponds to a geometrically thin, optically thick, steady accretion disk. The power per unit area generated by such a disk rotating around a central object is given by \cite{page-thorne} and used also in the study of disks around gravastars \cite{Harko2009c} and naked singularities \cite{Harko2010} reads $D(r) = \frac{\dot{M}}{4\pi r}\frac{\alpha}{a}\left(-\frac{d\Omega}{dr}
\right)
\frac{1}{(E-\Omega L)^2} \int^{r}_{r_{i}}(E-\Omega L)\frac{dL}{dr}dr$, where $\dot{M}$ is the accretion mass rate, $E,~L,~\Omega$ are the energy per unit mass, angular momentum per unit mass and angular velocity of test particles the disk is made of and $r_{i}$ is the inner edge of the disk. For black holes this radius is assumed to be at the ISCO ($r=6M$) of the hole.
For BSs we choose $r_i=0$ based on two considerations: i) BSs allow circular orbits in the whole spatial domain and thus there is no kinematical restriction to choose  $r_{i} = 0$ and ii) it has been shown that the luminosity of the disk for BSs considering this inner radius never reaches the value of Eddington luminosity in the whole spatial domain even for high accretion rates and thus there are no radiation pressure effects imposing a restriction on the inner edge location \cite{diego-acc}. Now, assuming it is possible to define a
local temperature the Steffan-Boltzmann law is used so that $D(r) = \sigma
T^4$, where $\sigma=5.67 \times 10^{-5}~ erg~s^{-1}~cm^{-2}~K^{-4}$ is the
Steffan-Boltzmann constant. Now, considering the disk emits as a black
body, we use the dependence of $T$ on the radial coordinate
and therefore the luminosity $L(\nu)$ of the disk for each frequency $\nu$ and the flux $F(\nu)$ can be calculated 
using the expression for the black body spectral distribution:
$L(\nu) = 4\pi d^2 F(\nu) =
\frac{16 \pi h}{c^2} \cos (\vartheta) \nu^3 \int^{r_f}_{r_i}
\frac{rdr}{e^{h\nu/kT} - 1}$, where $d$ is the distance to the source, $r_i$ and $r_f$
indicate the location of the inner and outer edges of the disk, $h=6.6256
\times 10^{-27}~ erg~s$ is the Planck constant, $k = 1.3805 \times
10^{-16}~ erg~K^{-1}$ is the Boltzmann constant and $\vartheta$ is the
disk inclination. The algorithm to construct the emission spectrum for
such a model of accretion disk around a BS and a BH is as follows: 1) Define the space-time functions $a$ and $\alpha$ by choosing one of the equilibrium configurations in Fig. 1a and
calculate $M$; 
2) Define the metric of the equivalent BH through $\alpha_{BH} =
\sqrt{1-2M/r}$ and $a_{BH} = 1/\sqrt{1-2M/r}$; 
3) Calculate the angular velocity, angular momentum and energy of a test
particle for both space-times $\Omega_{BS, BH}$, $L_{BS,BH}$,
$E_{BS,BH}$; 
4) Use such quantities to calculate the power emitted in both cases
$D_{BS}(r)$ and $D_{BH}(r)$; 
5) Calculate the temperature of the disk in both cases
$T_{BS}(r) = (D_{BS}(r)/\sigma)^{1/4}$ and
$T_{BH}(r) = (D_{BH}(r)/\sigma)^{1/4}$; 
6) Use such temperature to integrate the luminosity 
$L_{BS}(\nu)$ and $L_{BH}(\nu)$ for several values of $\nu$. 


\section{Mimicker configurations}

The free parameters of BSs that one can play with in order to produce a spectrum that mimics that due to the presence of a black hole are $m$ and $\Lambda$. As an example in Fig. 1b the power spectrum of a black hole is compared with two spectra due to two different BS configurations of the same mass but with different self-interaction $\Lambda$, one of which is a mimicker. It can be noticed how the spectrum due to one of the BSs lies on top of that due to the black hole. This is the effect that a black hole mimicker produces. For each value of $\Lambda$ it is possible to find mimickers by changing $m$. For instance, in Fig. 1a stars indicate the configurations corresponding to black hole mimickers for various values of $\Lambda$.

\begin{figure}[h]
\includegraphics[width=7cm]{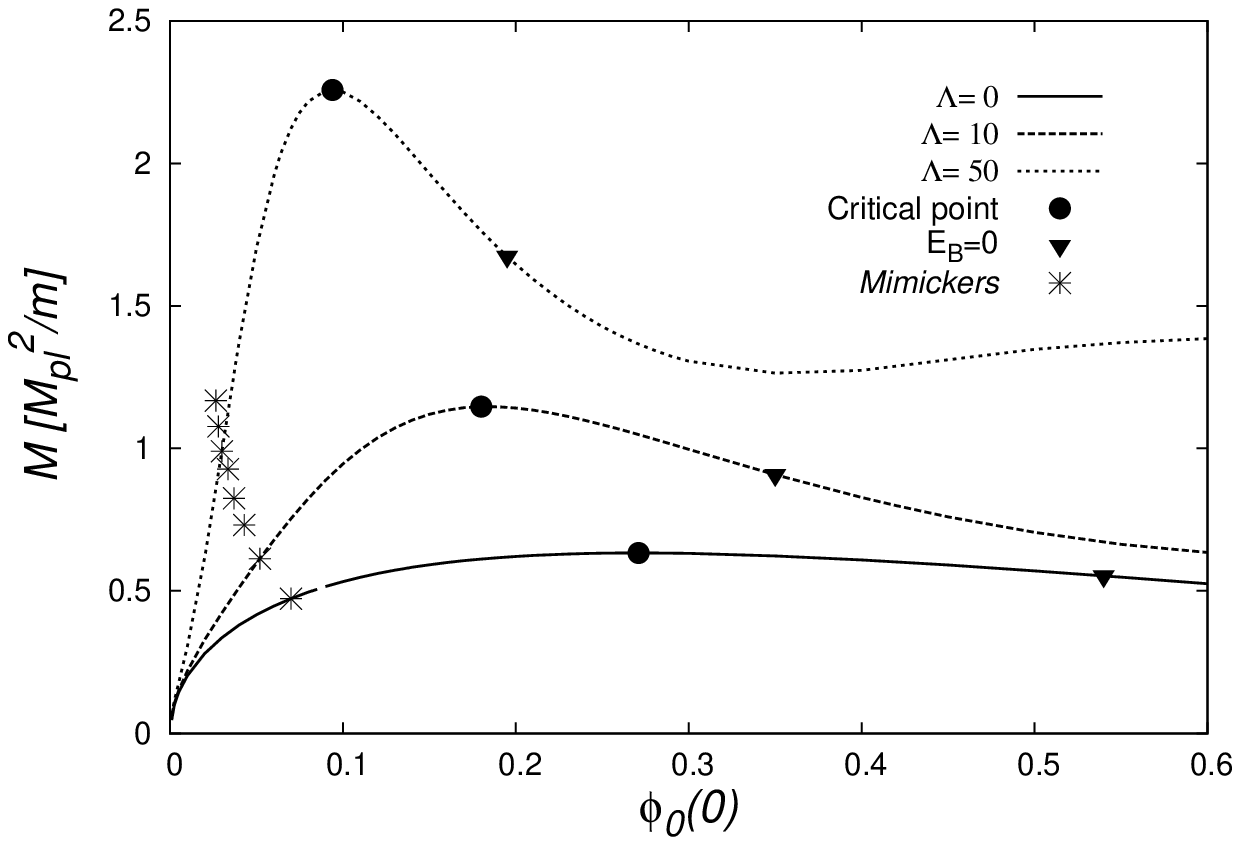}
\includegraphics[width=7.4cm]{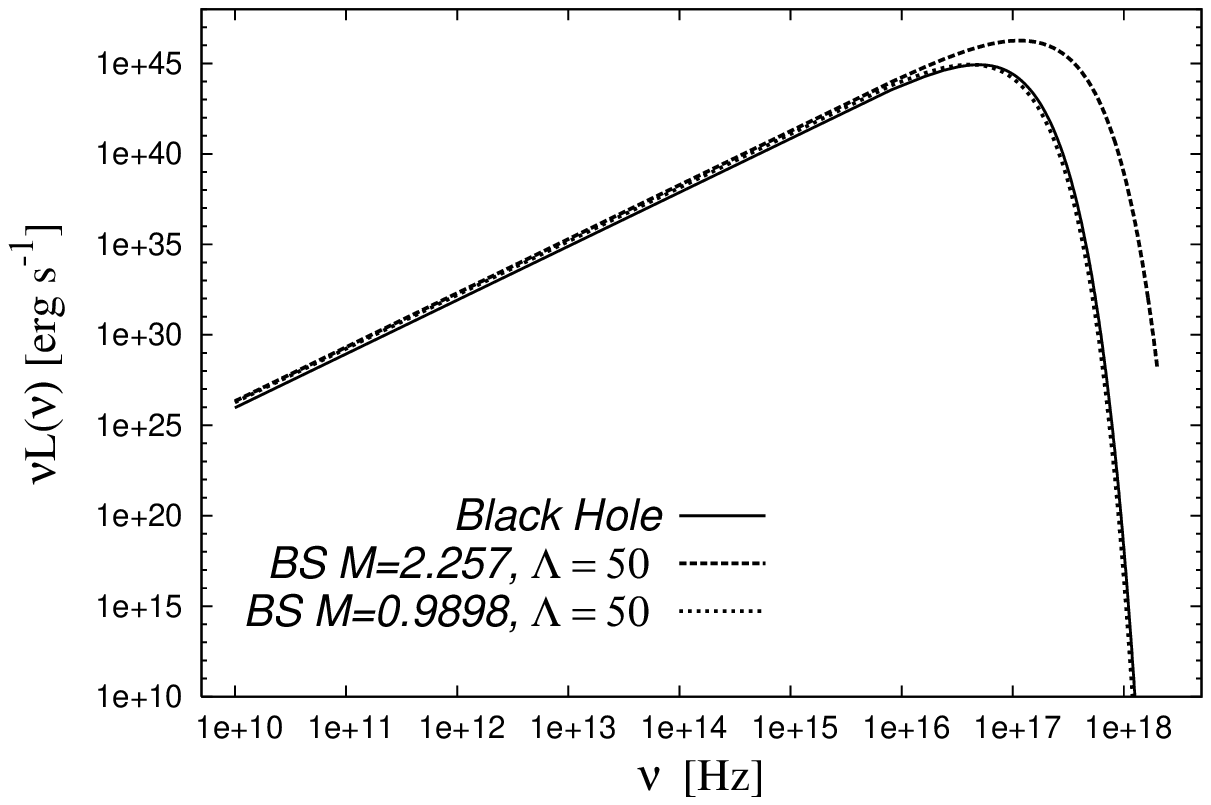}
\caption{\label{fig:mimickers} In the left panel it is shown a set of BSs configurations; each point of each curve corresponds to a BS configuration. Those configurations to the left of the maxima are stable, those between the maxima and the  The mimicker configurations are indicated with stars for values of $\Lambda=0,10,20,30,40,50,60,70$. It is expected that for  every value of $\Lambda$ there is a mimicker configuration that lies near the ones indicated for the values shown here. We show the mass of the inverted triangles are unstable with negative binding energy and collapse into black holes, whereas those to the right of the inverted triangles are unstable configurations that instead of collapsing explode \cite{Guzman2009RMF}. In the right panel we show an example of spectrum generated by a disk around  BS and around a Schwarzschild black hole.
The mimicker configuration used corresponds to a BS configuration with $M=0.9898M_{pl}^{2}/m$, $\Lambda=50$ and boson mass $m=5.25\times 10^{-22}$GeV, and works as a mimicker of a black hole with mass $M=10M_{\odot}$ with an accretion rate parameter $2\times 10^{-12}M_{\odot}/yr$. The spectra due to the black hole and to the mimicker are on top of each other, and for reference the spectrum due to a non-mimicker configuration is also shown.}
\end{figure}


\section{Discussion}

A summary of boson stars as black hole mimickers has been presented. It has been found that it is possible to find a boson star configuration that may supplant a black hole based on the emission spectrum of an accretion disk model. Nevertheless there are items that have to be addressed if boson stars are to still be considered black hole mimickers. The most important one is the question of whether or not boson stars are capable of trapping baryons, which eventually would form a mechanical surface that would behave like that of a baryonic compact object as assumed in studies of mimickers \cite{narayan}. The fact that boson star space-times are regular everywhere indicates that particles (e.g.  baryons) following geodesics would truly be accreted and retained on top of the boson star space-time; this would imply that there is nothing like accretion of matter, and instead, that the region of the boson star is a place where particles fall into and are expelled out, because it has been shown \cite{Guzman2006} that particles following radial trajectories would cross the origin and get out, and when test particles have a non-zero impact parameter there is an infinite potential barrier that forbids test particles to remain near the center of the boson star. However this is not a definitive answer, instead it is necessary to consider the self-gravity of baryons and their contribution to the geometry. With that information at hand it will be possible to determine whether or not boson stars are still potential black hole mimickers and then proceed to other more stringent astrophysics related tests like gravitational lensing and gravitational radiation.\\

\noindent {\bf Acknowledgments.} This research is partly supported by grants:
CIC-UMSNH-4.9 and CONACyT 106466.

\end{document}